\documentclass[aps,twocolumn,prd,showpacs,nofootinbib]{revtex4}
\usepackage{amsmath}
\usepackage{graphicx}
\usepackage{dcolumn}
\usepackage{bm}
\usepackage{amssymb}
\usepackage{latexsym}
\usepackage{color}

\def\be{\begin{equation}}
\def\ee{\end{equation}}
\def\bea{\begin{eqnarray}}
\def\eea{\end{eqnarray}}

\begin{document}

\title{Curvaton with nonminimal derivative coupling to gravity}

\author{Kaixi Feng$^1$}
\email{fengkaixi10@mails.gucas.ac.cn}

\author{Taotao Qiu$^{2,3,4}$}
\email{xsjqiu@gmail.com}

\author{Yun-Song Piao$^1$}
\email{yspiao@ucas.ac.cn}

\vspace{16mm}

\affiliation{$1$ School of Physics, University of Chinese Academy
of Sciences, Beijing 100049, China \vspace{1mm}} \affiliation{$2$.
Leung Center for Cosmology and Particle Astrophysics National
Taiwan University, Taipei 106, Taiwan} \affiliation{$3$.
Department of Physics and Center for Theoretical Sciences,
National Taiwan University, Taipei 10617, Taiwan} \affiliation{$4$.
Institute of Astrophysics, Central China Normal University,
Wuhan 430079, China}

\begin{abstract}
We show a curvaton model, in which the curvaton has a nonminimal
derivative coupling to gravity. Thanks to such a coupling, we find
that the scale-invariance of the perturbations can be achieved for
arbitrary values of the equation-of-state of background, provided
that it is nearly a constant. We also discussed about tensor perturbations, the local
non-Gaussianities generated by the nonminimal derivative coupling
curvaton model, as well as the adiabatic perturbations which are
transferred from the field perturbations during the curvaton
decay.

\end{abstract}

\maketitle

\section{introduction}

There have been large amount of literatures discussing about the
curvaton mechanism \cite{Lyth:2001nq}, see also \cite{Mollerach,
Linde:1996gt, Enqvist:2001zp, Moroi:2001ct}. In the curvaton
mechanism, it is assumed that the perturbations generated by
inflaton itself is negligible, while the curvaton field, which is
another light scalar field besides inflaton, is obliged to
generate the right amount of the primordial perturbations.
Although perturbations generated in this way are isocurvature
ones, they can be transferred into adiabatic ones at the end of
inflation, either after curvaton dominates over the inflaton, or
after the curvaton decays and reaches equilibrium with the
decaying products of inflaton \cite{Lyth:2001nq}. Since the
perturbations are generated by the curvaton field, the form of
inflaton can be less constrained, and large non-Gaussianities are
also possible. See
\cite{Sasaki:2006kq,Allahverdi:2006dr,DBIcurvaton,runningfnl,Cai:2011zx,Qiu:2011cy,higgs,Li:2013hga}
about related works.

In the simplest curvaton case, the curvaton field is just a canonical field with negligible mass and interactions. However, it can be extended to more complicated models with arbitrary forms of Lagrangian or coupling terms. In this letter, we study a kind of curvaton with its kinetic term nonminimally coupled to the gravity. As will be explained in the next section, this kind of coupling has very salient feature of giving rise to scale-invariant power spectrum without knowing the evolution behavior of the universe. Moreover, similar to the Galileon models, this model can also get rid of ``ghost modes", even if the nonminimal coupling may violate the Null Energy Condition. We give full analysis of this model, especially on the perturbations. We study how it generates scale-invariant power spectrum for scalar perturbations, as well as what the tensor perturbations will be like. The stabilities of these perturbations gives further constraints on this model. We also study how the field perturbations can be transferred into curvature perturbations, as well as the local-type non-Gaussianities generated by this model.

This paper is organized as follows: in Sec. II we generally summarize our motivation of having nonminimal derivative coupling in the curvaton model. In Sec.III, we introduce the nonminimal derivative coupling curvaton, and briefly review its evolution behavior in background level. Sec. IV and Sec. V devotes themselves in the perturbation generated by the curvaton model. In Sec. IV, we analyze the linear scalar perturbation and its power spectrum. In the first subsection, we see that the power spectrum is indeed scale-invariant as we expected, although gravitational perturbations are considered for a comprehensive study, and small tilt of the spectrum can be given by the corrections from the potential term of the curvaton. In the second subsection we consider the tensor perturbations of our model. We find that the tensor perturbation gives more tight constraints on our model, and that a healthy tensor perturbation in our model can hardly be deviated from that of a minimal coupling single scalar model. In Sec. V we discuss about how the field perturbations can be transferred into adiabatic ones through the decay of curvaton or equilibrium with the background of the universe, and also derive the local non-Gaussianity generated by the curvaton in this section. We make our final conclusion in Sec. VI.

\section{Our Motivation}
In general, the curvaton is simply acted by a canonical scalar field which decouples from other matters and minimally couples to gravity. The action of the curvaton field $\varphi$ is described as \be \label{curvaton1} {\cal S}_\varphi=\int d^4x\sqrt{-g}P(X,\varphi)~,\ee where $P$ is an arbitrary function of $\varphi$ and its kinetic term: $X\equiv-\nabla_\mu\varphi\nabla^\mu\varphi/2$. \footnote{We adopt the notation of sign difference as $(-,+,+,+)$.} The simplest case is that $\varphi$ is a canonical field, thus $P(X,\varphi)$ reduces to the form of $X-V(\varphi)$. Setting $\varphi\rightarrow\varphi_0+\delta\varphi$, one can get the perturbed action of curvaton as: \be\label{perturb} \delta S_\varphi=\int d^3xd\eta\frac{a^2Q}{c_s^2}\left[{\delta\varphi^\prime}^2-c_s^2(\partial \delta\varphi)^2\right]~, \ee where \be Q\equiv P_{,X}~,~~~c_s^2\equiv\frac{P_{,X}}{\rho_{,X}}~,\ee and $^\prime$ means derivative with respect to conformal time $\eta$. Here we have assumed its effective mass is negligible. Current observational data favors the scale-invariance property of primordial perturbations \cite{Ade:2013zuv}. As can be derived from (\ref{perturb}), the scale-invariance of the perturbations generated by curvaton requires \cite{Qiu:2012ia} \be\label{z} z^2\equiv\frac{a^2Q}{c_s^2}\sim\frac{1}{(\eta_*-\eta)^2}~~~\text{or}~~~(\eta_*-\eta)^4~,\ee where the first case is for that the dominant mode of perturbations is a constant one, while the second is for that the dominant mode is an increasing one.

For the canonical field case, one have $Q=1$. Provided the universe evolves with a constant EoS $w$, one roughly has $a\sim-[H(\eta_*-\eta)]^{-1}$, thus from  Eq. (\ref{z}), it can only be that $H\sim\text{constant}$, which is inflation ($w\simeq-1$) for the first case, or $H\sim \eta^3$, which is the matter-like contraction ($w\simeq0$) for the second case. Here we also neglect the variation of $c_s^2$. Therefore we see that, the perturbations generated by curvaton is scale invariant only for limited cases where few values of $w$ can be chosen, which we think is too tight a constraint on the evolution behavior of the early universe.

Can we have a model of which Eq. (\ref{z}) is satisfied automatically, independent of how the universe evolves, whether expands or contracts, and whatever $w$ is? If the answer is yes, then we can have much wider possibilities of the early universe evolution, which can allow much more fruitful interesting cosmological phenomenons. To realize this, basically we need either $Q$ or $c_s^2$, or both, change with time. For example, if we still treat $c_s^2$ as a constant and only $Q$ to be time-varying, Eq. (\ref{z}) requires $Q$ scale as: \be\label{Q} Q\sim\frac{1}{a^2(\eta_*-\eta)^2}~~~\text{or}~~~(\eta_*-\eta)^4/a^2~\ee for constant/increasing-mode-dominating case, respectively.

There have been some possibilities of time-varying $Q$ presented in the literature, in which the curvaton field becomes non-canonical. The popular example recently proposed is to have curvaton generate scale-invariant perturbations via the so-called ``conformal mechanism" \cite{Libanov:2011bk,Hinterbichler:2011qk,Creminelli:2011mw}, which has been applied on Galileon-genesis \cite{Creminelli:2010ba} (see also slow expansion scenario for different case \cite{slow}) or Galileon bounce models \cite{Qiu:2011cy}. Clever idea as it is though, in these models $Q$ is written as functions of background field $\phi$, therefore depends severely on the details of the background evolution. In general, this will need more or less tuning of the background
field and become less controllable. The aim of this paper is to look for a kind of model of which the variation of $Q$ is universal, giving rise to required behavior (\ref{Q}) without worrying about the background evolution.

As has been mentioned before, the relation $a\sim-[H(\eta_*-\eta)]^{-1}$ holds universally for arbitrary evolution, provided that the EoS $w$ is a constant. Therefore, the first relation of (\ref{Q}) indicates that $Q$ should be proportional to $H^2$. Since $Q\sim P_{,X}$, a natural conjecture of the kinetic term of the lagrangian (\ref{curvaton1}) can be $RX$ or $G_{\mu\nu}\partial^\mu\varphi\partial^\nu\varphi$, where $R$ is the Ricci scalar and $G_{\mu\nu}$ is the Einstein tensor, respectively. Both of the two terms appears very commonly in the literature as ``nonminimal derivative coupling" \cite{Amendola:1993uh}. In homogeneous and isotropic FRW background, both of the two terms will be reduced to $H^2\dot\phi^2$, although the first one has a correction from time-derivative of $H$. From this property, we expect that models with such a kinetic term can automatically satisfy the relation (\ref{Q}), without knowing exactly the detailed background evolution of the universe. This is our very motivation of studying this kind of curvaton model in this paper.

\section{the model}
The action of nonminimal derivative coupling curvaton is considered as \be\label{action} \mathcal{S}=\int\mathrm{d}^4x\sqrt{g}\Big[\frac{R}{16\pi G}+\frac{\xi}{M^2}G_{\mu\nu}\partial^\mu\varphi\partial^\nu\varphi-V(\varphi)+{\cal L}_{bg}\Big]~,\ee where $G_{\mu\nu}$ is the Einstein tensor: $G_{\mu\nu}\equiv R_{\mu\nu}-g_{\mu\nu}R/2$, and $\xi$ is an arbitrary coefficient.

Being first proposed by Amendola in 1992 \cite{Amendola:1993uh} where the most general terms was given, nonminimal derivative coupling has been applied to various aspects of cosmology, for example, see \cite{Capozziello:1999xt,Granda:2011zk,Germani:2010gm} for inflation (see \cite{Sadjadi:2012zp} for the reheating process), see \cite{Capozziello:1999uwa,Granda:2009fh,Saridakis:2010mf,Granda:2011eh,Sadjadi:2010bz,Banijamali:2011qb} for dark energy, 
see \cite{Banijamali:2012kq} for bouncing cosmology, and see
\cite{Chen:2010ru,Lin:2011zzd,Rinaldi:2012vy} for the study of
black hole physics making use of non-minimal derivative couplings.
In \cite{Cartier:2001is} (see also \cite{Sushkov:2009hk}), it was
pointed out that such term only leads to second order field
equations and showed the exact cosmological solutions, in
\cite{Daniel:2007kk} Daniel and Caldwell analyzed the
(in)stabilities of and put constraints on such kind of model, and
in \cite{Gao:2010vr}, Gao shows that nonminimal derivative
coupling field can act not only as dark energy, but also as dark
matter. Moreover, when coupled to several Einstein tensors, it can
also gives rise to inflationary behavior.

The nonminimal derivative coupling can also be viewed as a subset
of Galileon models \cite{Nicolis:2008in,Deffayet:2009mn}. One of
the appealing properties of this kind of model is that, while the
action contains higher derivative of the field or nonminimal
coupling, due to the delicate design of the lagrangian, the
equation of motion of the field remains of second order, which can
be free of ghost. The simplest Galileon model, which contains the
coupling term $(X\Box\varphi)$, can be applied to curvaton
scenario, where the higher derivative term can give rise to local
non-Gaussianities of ${\cal O}(10)$ \cite{Wang:2011dt}. In this
paper we try to apply another kind of Galileon model on the
curvaton scenario.

From action (\ref{action}), one can straightforwardly obtain the energy density and pressure are expressed as \bea\label{rho} \rho_\varphi&=&\frac{9\xi}{M^2}H^2\dot\varphi^2+V(\varphi)~,\\ \label{p} P_\varphi&=&-\frac{\xi}{M^2}(3H^2\dot\varphi^2+2\dot H\dot\varphi^2+4H\dot\varphi\ddot\varphi)-V(\varphi)~,\eea respectively. Moreover, the equation of motion (EoM) of $\varphi$ can be written as: \be\label{eom} \frac{6\xi}{M^{2}}H^{2}\ddot{\varphi}+\frac{6\xi}{M^{2}}(2\dot{H}+3H^{2})H\dot{\varphi}+V_{\varphi}=0~.\ee It's also useful to define parameters like: \be\label{para} y\equiv\frac{\xi}{M^{2}}\dot{\varphi}^{2}~,~\eta\equiv\frac{\ddot{\varphi}}{H\dot\varphi}~,~\epsilon\equiv-\frac{\dot{H}}{H^{2}}~,~\epsilon_{\phi}\equiv\frac{\dot{\phi}^{2}}{M_{p}^{2}H^{2}}~,\ee which we will use later. One can also notice that $\dot{y}=2\frac{\xi}{M^{2}}\dot{\varphi}\ddot{\varphi}=2Hy\eta$.

As a general study, in the following we will briefly review the evolution behavior of the curvaton field $\varphi$ according to Eqs. (\ref{rho}-\ref{eom}). We will divide the whole analysis into three classes: \\
{\it 1) $V(\varphi)=0$.} In this class, the energy density of $\rho_\varphi$ is contributed only by its kinetic term, and the EoM (\ref{eom}) can be easily solved without $V_\varphi$. According to the solution, $\dot\varphi$ scales as $a^{-3}H^{-2}$, then the energy density scales as $\rho_\varphi\sim H^2\dot\varphi^2\sim a^{-6}H^{-2}$. The scaling behavior of $a$ and $H$ are determined by the background energy density $\rho_{bg}\propto a^{-3(1+w)}$, so one can straightforwardly have $\rho_{bg}\sim\text{constant.},\rho_\varphi\propto a^{-6}$ for $w=-1$, $\rho_{bg}\propto a^{-3},\rho_\varphi\propto a^{-3}$ for $w=0$, $\rho_{bg}\propto a^{-4},\rho_\varphi\propto a^{-2}$ for $w=1/3$, and $\rho_{bg}\propto a^{-6},\rho_\varphi\sim\text{constant.}$ for $w=1$. One can see that this forms a duality between the energy densities of the background and $\varphi$, the relation of which is $\rho_\varphi\rho_{bg}\sim a^{-6}\sim\rho_m^2$, where $\rho_m$ denotes energy density of non-relativistic matter. This is an interesting property of the nonminimal derivative coupling models subdominant in the universe. Similar arguments has also been made in  \cite{Gao:2010vr}. \\
{\it 2) $V(\varphi)\neq0$, $V_\varphi=0$.} In this case, the EoM of $\varphi$ is the same as in the above case, so $\varphi$ has the same solution as $\dot\varphi\sim a^{-3}H^{-2}$, however $\rho_\varphi$ will be added by a constant potential $V=V_0$. For power-law ansatz solution, $a(t)\sim t^{2/3(1+w)}$, one can get the scaling of the kinetic term of $\rho_\varphi$, $\xi H^2\dot\varphi^2\sim a^{3(w-1)}$. For different background where $w$ and scaling of $a$ is different, this term can be increasing or decreasing, or remains constant, and $\rho_\varphi$ can be dominated by either kinetic term or potential. In the expanding universe, when $w>1$ the kinetic term is increasing and will dominate $\rho_\varphi$, and when $w<1$ it is decreasing and $\rho_\varphi$ will be dominated by the potential. Things become vice versa in contracting universe, and in both cases, the two part will scale in-phase and contribute to $\rho_\varphi$ together for $w=1$. \\
{\it 3) $V(\varphi)\neq0$, $V_\varphi\neq0$.} This is the most general and complicated case and usually the exact solution cannot be obtained analytically. Even though, for some simple cases, we can still find some ansatz solutions. One of the example is the scaling solution, of which all the terms in EoM and Friedmann equations has the same scaling w.r.t $t$. Assuming $\varphi\sim t^c$ where $c$ is a constant parameter, one finds that to have the terms in EoM have the same scaling requires $V_\varphi\sim t^{c-4}$, which in turn gives $V(\varphi)\sim\varphi^{2-4/c}$. Thus one has $\rho_\varphi\sim t^{2c-4}$, while $\rho_{bg}\sim H^2\sim t^{-2}$. For $c<1$, when $|t|\rightarrow\infty$ (late time or early time in bouncing cosmology), $\rho_\varphi$ will not affect the background much, but will have significant effect at $|t|\rightarrow 0$ (approaching to Big-Bang Singularity). Vice Versa for $c>1$. Note that for $c=1$ where $\dot\varphi$ becomes constant, the energy density of $\varphi$ will have the same scaling as that of the background.

\section{perturbation}
\subsection{scalar perturbation}
In Sec. II, we demonstrated that this kind of curvaton model can give rise to scale-invariant power spectrum without knowing the exact behavior of the universe, which is due to the nonminimal derivative coupling. There is, though, a small stumbling block in front of us before we cheer for such an easy but interesting expectation. As we introduce the nonminimal coupling of curvaton to gravity, the gravitational perturbations, which has always been neglected for curvaton models, might get invoked, although one of the components can be gauged away. Such an effect may or may not change the perturbed action (\ref{perturb}) substantially, which is the basis of our result, so one should be careful in treating perturbations of such a model and consider the gravitational perturbations as well. So in this section, we will analyze the perturbations of this model with careful calculation. We will show that, fortunately, the gravitational perturbations will indeed not play much role in our model, and our expected result still holds very well.

The perturbed metric can be written as follows:
\be\label{adm}
ds^{2}=-N^{2}dt^{2}+h_{ij}(dx^{i}+N^{i}dt)(dx^{j}+N^{j}dt)~,\ee
where $N$ is the lapse function, $N^i$ is the shift vector, and $h_{ij}$ is the induced 3-metric. One can then perturb these functions as:
\be
N=1+\alpha~,~N_i=\partial_i\beta~,~h_{ij}=a^2(t)e^{2\psi}\delta_{ij}~,\ee
where $\alpha$, $\beta$ and $\psi$ are the scalar metric perturbations. As for curvaton models, it is convenient to consider the spatial-flat gauge, where $\psi=0$. Moreover, the pertubation of $\varphi$ field is
\be
\varphi\rightarrow \varphi(t)+\delta\varphi(t,\bf{x})~.\ee
The perturbation generated by the background field $\phi$ is also neglected.

Using this, we can expand our action (\ref{action}) up to the second order. The expansion and reduction process are rather tedious, as can be seen for the nonminimal derivative coupling term in Appendix, after which we can obtain the constraining degrees of freedom $\alpha$ and $\beta$, which appears to be:
\be
\alpha=a_1\dot{\delta\varphi}+a_2\delta\varphi~,~~~\partial^2\beta=b_1\dot{\delta\varphi}+b_2\delta\varphi+b_3\partial^2\delta\varphi
~,\ee
where we define
\bea
a_1&\equiv&-\frac{2\xi\dot{\varphi}/M^2}{M_p^2-3y}~,~a_2\equiv\frac{3\xi H\dot{\varphi}/M^2}{M_p^2-3y}~,\\
b_1&\equiv&a^{2}\frac{[a_{1}\frac{\dot{\phi}^{2}}{2H}-\frac{9\xi H\dot{\varphi}}{M^{2}}-3H a_{1}(M_{p}^{2}-6y)]}{M_p^2-3y}~,\\
b_2&\equiv&a^{2}\frac{[a_{2}\frac{\dot{\phi}^{2}}{2H}-3H a_{2}(M_{p}^{2}-6y)-\frac{V_{,\varphi}}{2H}]}{M_p^2-3y}~,\\
b_3&\equiv&-\frac{2\xi\dot{\varphi}/M^{2}}{M_{p}^{2}-3y}~.\eea

Substituting this into (\ref{action}), and use conformal time $\eta$ instead of cosmic time $t$, one will finally find the second order perturbation action appears as: \be\label{perturb2} \delta S_\varphi=\int d\eta  d^{3}xa^2\frac{Q}{c_s^2}\Big[{\delta\varphi}^{\prime 2}-c_s^2\partial_{i}\delta\varphi\partial^{i}\delta\varphi-\frac{1}{2}\frac{a^2c_s^2m_{eff}^{2}}{Q}\delta\varphi^{2}\Big]~,\ee where\bea \label{Qcs2meff}
Q&\equiv&-\frac{\xi H^{2}}{M^{2}}\Big[\frac{4y(4-\epsilon+2\eta)}{M_{p}^{2}-3y}+\frac{24y^{2}\eta}{(M_{p}^{2}-3y)^{2}}+2\epsilon-7\Big]~,\nonumber\\ c_s^2&\equiv&\frac{Q M^{2}}{\xi H^{2}}\Big[2y\frac{(12+\epsilon_{\phi})M_{p}^{2}-18y}{(M_{p}^{2}-3y)^{2}}+3\Big]^{-1}~,\nonumber\\
m_{eff}^{2}&=&\frac{\xi}{M^{2}}\frac{3M_{p}^{2}H^{4}y}{(M_{p}^{2}-3y)^{2}}[3(6\epsilon-3\epsilon_{\phi}-4\eta\frac{M_{p}^{2}+3y}{M_{p}^{2}-3y})\nonumber\\ &&-2\epsilon_{\phi}(\frac{\dot{\epsilon}_{\phi}}{H\epsilon_{\phi}}-3\epsilon+2\eta\frac{M_{p}^{2}+3y}{M_{p}^{2}-3y})]+\frac{M_{p}^{2}-y}{M_{p}^{2}-3y}V_{\varphi\varphi}\nonumber\\ &&-54\frac{\xi}{M^{2}}\frac{H^{4}y^{2}}{(M_{p}^{2}+3y)^{2}}(9-3\epsilon+4\eta\frac{M_{p}^{2}}{M_{p}^{2}-3y})\nonumber\\ &&-12\frac{\xi}{M^{2}}\frac{H^{4}y}{M_{p}^{2}-3y}(6+\eta\frac{M_{p}^{2}+3y}{M_{p}^{2}-3y})(\eta-2\epsilon+3)~.\eea Note that in the above formulation, we have made use of the background equation of motion (\ref{eom}) as well as the definitions of the parameters in Eq. (\ref{para}).

Define $z\equiv a\sqrt{Q}/c_s$, we can get the equation of motion of $\delta\varphi$ in its neat form as:
\be
(z\delta\varphi)^{\prime\prime}+(c_s^2k^2-\frac{z^{\prime\prime}}{z}+\frac{1}{2}\frac{a^4m_{eff}^2}{z^2})(z\delta\varphi)=0~,\ee and as will be shown explicitly later, for the case of $Q\sim H^2$ and negligible $m_{eff}^2$, we have the approximate solution
\be\label{solvarphi}
\delta\varphi=k^\nu(\eta_*-\eta)^{\nu+\frac{3}{2}}~,~~~k^{-\nu}(\eta_*-\eta)^{\nu+\frac{3}{2}}~,~\nu\simeq \frac{3}{2}~,\ee where we've made use of $z^2\sim a^2H^2\sim (\eta_*-\eta)^{-2}$. From these solution we can see that, whether expanding or contracting the universe will be, the last mode (which is constant) will dominate over the first one (which is decaying), and gives rise to scale-invariant power spectrum. This is because that the factor $Q$ has some ``faking" effect, making the perturbation $\delta\varphi$ ``feel" itself in a de-Sitter expanding phase, even though the real evolution is not. This is an interesting application of the ``conformal mechanism" in \cite{Libanov:2011bk,Hinterbichler:2011qk,Creminelli:2011mw} and is the essential property of this model that we are pursuing. Moreover, the power spectrum of $\delta\varphi$, ${\cal P}_{\delta\varphi}$, and the spectral index, $n_s$, are defined as:
\be\label{spectrum}
{\cal P}_{\delta\varphi}\equiv\frac{k^3}{2\pi^2}\frac{|\delta\varphi|^2}{M_p^2}~,~n_s-1\equiv\frac{d\ln{\cal P}}{d\ln k}~,\ee respectively.

To have analytical solutions we take two interesting limits. The first case is that $|y|\ll M_{p}^{2}$, meaning that the velocity of the curvaton field is quite slow and the kinetic term of curvaton is negligible. In this case, we have:
\bea\label{smally}
Q\simeq-\frac{\xi H^{2}}{M^{2}}(2\epsilon-7)~,~c_{s}^{2}\simeq-\frac{1}{3}(2\epsilon-7)~,~m_{eff}^{2}\simeq V_{\varphi\varphi}~.\eea
When $V(\varphi)$ is sufficiently flat, $V_{\varphi\varphi}\ll H^2$, the effective mass term can also be neglected.
In order to make this model free of ghost and gradient instabilities, we require $Q>0$, $c_s^2>0$, which leads to
\be\label{constraint1} \xi>0~,~~~w=-1+\frac{2}{3}\epsilon<\frac{4}{3}~,\ee which is the region of viability of our model in this case.

One can get the power spectrum of $\delta\varphi$ from Eqs. (\ref{solvarphi}) and (\ref{spectrum}), which is:
\be
{\cal P}_{\delta\varphi}=\frac{H^2}{4\pi^2M_p^2c_s^3Q}=\sqrt{\frac{27}{(7-2\epsilon)^5}}\frac{M^{2}}{4\pi^2M_p^2\xi}~,\ee where in the last step, we've made use of the results in Eq. (\ref{smally}). Note also that our result indicates that the spectrum of $\delta\varphi$ is determined by the cutoff scale $M$ instead of $H$, and $H$ will not be constrained by the spectrum. However, as will be seen very soon, the dependence on $H$ will be turned on when the perturbations of curvaton field transfer into the adiabatic perturbations.

In this result, the power spectrum is exactly scale-invariant if there is no other correction term, however, the observational data from PLANCK \cite{Ade:2013zuv} favors small tilt in the power spectrum index. Recall that we have turned off the mass term for simplicity, and when this term is turned on, we may get this small correction. Considering mass term from (\ref{smally}), the equation of motion becomes:
\be\label{eompert}
(z\delta\varphi)^{\prime\prime}+(c_s^2k^2-\frac{z^{\prime\prime}}{z}+\frac{1}{2}\frac{a^4V_{\varphi\varphi}}{z^2})(z\delta\varphi)=0~.\ee
For a specific choice, we choose $V_{\varphi\varphi}\sim H^4\sim t^{-4}$, which can be reconstructed to give $V(\phi)\sim t^{2c-4}\sim \varphi^{2-4/c}$, according to our background analysis in Sec. II. This ansatz gives the scaling of the last term w.r.t. the conformal time, $a^4V_{\varphi\varphi}/2z^2\propto (\eta_*-\eta)^{-2}$. Setting the prefactor to be $\Delta_1$, and moreover since $z^{\prime\prime}/z\sim 2/(\eta_*-\eta)^2$, Eq. (\ref{eompert}) becomes
\be
(z\delta\varphi)^{\prime\prime}+[c_s^2k^2-\frac{2-\Delta_1}{(\eta_*-\eta)^{2}}](z\delta\varphi)=0~,\ee which one can solve to get the spectral index $n_s$ (defined through $n_s\equiv d\ln{P}/d\ln k$) of our model in this case:
\be\label{nsphi1} n_s-1=\frac{2}{3}\Delta_1~.\ee

The second case of the analytical solution of our model is that $|y|\gg M_p^2$, indicating that the curvaton moves with a very large speed. In realistic models, this case is somehow dangerous, since this might give a large kinetic term to curvaton, and to make the energy density of the curvaton field exceed that of the background, one might need also a large potential term with opposite sign to cancel the kinetic energy, which requires fine-tuning in some level, so one may need to be careful to make it a healthy model. However here as a complete study, we will also consider this case. From Eq. (\ref{Qcs2meff}) we have:
\bea\label{largey}
Q&\simeq&-\frac{\xi H^{2}}{3M^{2}}(10\epsilon-37)~,~c_{s}^{2}\simeq\frac{1}{3}(10\epsilon-37)~,\nonumber\\ ~m_{eff}^{2}&\simeq&-2\frac{\xi}{M^{2}}H^{4}(9+15\epsilon-6\eta+2\eta^{2}-4\eta\epsilon)\nonumber\\ &&+\frac{1}{3}V_{\varphi\varphi}~,\eea
and the positivity of $Q$ and $c_s^2$ requires
\be\label{constraint2} \xi<0~,~~~w>\frac{22}{15}~.\ee
Different from the previous case, here we obtained a correction that is proportional to $H^4$ which is brought by considering metric perturbation. Similarly, one can also obtain corrections in the equation of motion of $\delta\varphi$, which is proportional to $(\eta_*-\eta)^{-2}$. Setting the prefactor to be $\Delta_2$, one can get the power spectrum and its index as:
\bea
{\cal P}_{\delta\varphi}&=&\frac{H^2}{4\pi^2M_p^2c_s^3Q}\nonumber\\ &=&\sqrt{\frac{243}{(10\epsilon-37)^5}}\frac{M^{2}}{4\pi^2M_p^2(-\xi)}~,\\
\label{nsphi2} n_s-1&=&\frac{2}{3}\Delta_2~.\eea

As a side remark, we should mention that a non-vanishing $V_{\varphi\varphi}$ can also provide tilt of the spectrum as well. However, as can be seen in next section, when we consider the tensor perturbations of this model, this case will be ruled out since it induces the instability of tensor perturbations by have the sound speed of gravitational waves $c_T^2<0$.

\subsection{tensor perturbation}
The recently release PLANCK data not only did a well measurement
for scalar perturbations in the early universe, but also plans to
measure the tensor perturbations, i.e. the gravitational waves.
Whatever the future result is, one conclusion that can now be
confirmed is that the signature of gravitational waves is quite
small. This can already have some constraints on theoretical
models. To make our analysis complete, we also consider the tensor
perturbations of our model. \footnote{We thank the referee for
pointing us this issue.} The metric containing tensor perturbation
is written as: \be\label{admtensor}
ds^{2}=-dt^{2}+a^2(t)(\delta_{ij}+\gamma_{ij})dx^{i}dx^{j}~,\ee
where $\gamma_{ij}(x)$ is the tensor perturbation satisfying
traceless and transverse conditions: \be
\gamma^i_i=0~,~~~\partial_i\gamma^i_j=0~.\ee Note that in our
model, the field part also have contributions to tensor
perturbations due to the nonminimal coupling, we can perturb the
action (\ref{action}) up to second order as:
\be\label{perturbtensor} \delta S_{\varphi}^T = \frac{1}{8}\int dt
d^3x a^3[{\cal F}\dot\gamma_{ij}^{2}-{\cal
G}\frac{(\nabla\gamma_{ij})^2}{a^2}]~,\ee where we defined \be
{\cal F}=M_p^2-y~,~~~{\cal G}=M_p^2+y~,\ee where $y$ is defined as
in (\ref{para}). Note that this action is in consistent with that
of Generalized Galileon action in \cite{Kobayashi:2011nu}. From
this action, one can read that the squared sound speed of
gravitational waves is: \be c_T^2\equiv\frac{{\cal G}}{{\cal
F}}=\frac{M_p^2+y}{M_p^2-y}~.\ee

In the previous analysis for scalar perturbation, we discussed
about two limited cases, namely $|y|\ll M_p^2$ and $|y|\gg M_p^2$.
However, from the above expression of $c_T^2$ we can see that, the
last case will induce $c_T^2\simeq-1<0$, which will induce an
unstable tensor perturbation. So this case should be abandoned,
leaving only the first case, which gives $c_T^2\simeq 1$, namely a
healthy tensor perturbation. Therefore in the following, we will
only consider this case. In this case, we have ${\cal F}\simeq
M_p^2$, ${\cal G}\simeq M_p^2$, and thus the behavior of the
tensor perturbation will actually be very close to that in the
case of minimal coupling single scalar, in which the primordial
tensor perturbation spectra from various expanding and contracting
phases has been calculated in \cite{Piao:2004jg}. Taking the
conditions of constant $w$, one can derive the equation of motion
for $\gamma_{ij}$ from Eq. (\ref{perturbtensor}) and get the
solution: \be
\gamma_{ij}=\text{constant.},~~~~\int\frac{dt}{a^3(t)M_p^2}~,\ee
where $a(t)$ can be parametrized as $a(t)\sim t^{2/3(1+w)}$. From
this result we can easily see that the tensor perturbation is
dominated by its constant mode when $w>1$ for contracting phase or
$w>-1/3$ for expanding phase where the varying mode is actually
decreasing, while by its varying mode when $-1/3<w<1$ for the
contracting phase where the varying mode is actually growing
\footnote{The boundary of $-1/3$ is set for the requirement of
avoidance of the horizon problem.}. By detailed calculation, the
tensor spectrum is obtained as: \be {\cal P}_T\sim
k^{3}|\gamma_{ij}|^2\sim
\frac{H^2}{M_p^2}\big(\frac{k}{k_0}\big)^{n_T}~, \ee where the
spectral index \bea
n_T&=&\frac{6(1+w)}{1+3w}~~~~(w>1)~,\nonumber\\
&or&\frac{12w}{1+3w}~~~~(-\frac{1}{3}<w<1)~\eea for contracting
phase and \be n_T=\frac{6(1+w)}{1+3w}~~~~(w<-\frac{1}{3})~\ee for
expanding phase. Here $k_0$ denotes some pivot wavenumber.

One can see from the result that, in order not to get too much
gravitational waves, we need either scale-invariant tensor
spectrum with the amplitude $|{\cal P}_T|^{1/2}\sim M_p^{-1}H\sim
10^{-5}$, or blue-tilted tensor spectrum of which the amplitude
can be larger (namely, $H$ can be larger than $10^{-5}M_p$), which
could be suppressed by power-laws of $k$ on large scales (cf.
\cite{Qiu:2011cy}). This requires the background equation of state
be either no less than 0 (for contracting phase) or no more than
$-1$ (for expanding phase) \footnote{We do not mean that $w$'s out
of this range are completely ruled out, since this is only a
theoretical estimation. For example, as in usual inflation case,
although $w>-1$ causes a red-tilted tensor spectrum which will be
raised on large scales, it can still survive if the deviation of
$w$ from $-1$ is not too much (namely under slow-roll condition),
such that the raising effect does not conflict with the
observational data.}. Combining the constraints on $w$ that was
obtained previously for scalar perturbations, we can conclude that
the viable condition for our model is that \bea \left\{
\begin{array}{l} 0<w<\frac{4}{3}~~~~{\rm for}\,\,\,{\rm
contracting}\,\,\, {\rm phase},\\\\ w<-1~~~~{\rm for}\,\,\,{\rm
expanding}\,\,\,
{\rm phase},\\
\end{array}\right.\eea which is another important conclusion of our model.

\section{the creation of curvature perturbation and Local-type non-Gaussianity}
In the above section, we showed that our model is able to give rise to scale-invariant perturbations with large range of universe evolution, provided constraints (\ref{constraint1}) and (\ref{constraint2}) holds in order to keep the perturbation of the model well-defined. However, these perturbations are of isocurvature ones. The adiabatic curvature perturbation, which can be observable, can be obtained in two ways: one is when the background decays and the curvaton dominates the universe, and the other is the curvaton and background decay simultaneously and their decay products become equilibrium. Here we assume that the background decays into relativistic matter for simplicity. The final curvature perturbation can be expressed as:
\be\label{zeta}
\zeta=\frac{\delta\rho_\varphi}{4\rho_r+3(\rho_\varphi+P_\varphi)}~,
\ee where the density perturbation $\delta\rho_\varphi$ can furtherly be expanded with respect to $\delta\varphi$. The linear and next-to-linear order of $\delta\rho_\varphi$ is given by:
\bea
\delta^{(1)}\rho_\varphi&\equiv&\rho_{\varphi,\varphi}\delta\varphi\nonumber\\&\simeq&\Big(-\frac{18\xi\epsilon H^3\dot\varphi}{M^2}+V_{,\varphi}\Big)\delta\varphi~,\\
\delta^{(2)}\rho_\varphi&\equiv&\frac{1}{2}\rho_{\varphi,\varphi\varphi}\delta\varphi^2 \nonumber\\
\label{deltarho2}&\simeq&\frac{1}{2}\Big(\frac{54\xi\epsilon^2H^4}{M^2}+V_{,\varphi\varphi}\Big)\delta\varphi^2~, \eea respectively.

Now we can consider the two ways separately. If the curvaton dominates the universe, say $\rho_\varphi\gg\rho_r$, from Eq. (\ref{zeta}) we have:
\bea
\zeta^{A} &\simeq&
\frac{\delta^{(1)}\rho_\varphi}{3(\rho_\varphi+P_\varphi)}
\simeq\frac{\rho_{\varphi,\varphi}}{3(\rho_\varphi+P_\varphi)}\delta\varphi
\nonumber\\
\label{zetaa}&\simeq&-\frac{3\epsilon H_*}{\dot\varphi_*
(3+\epsilon-2\eta)}\delta\varphi~, \eea where $H_*$ and $\dot\varphi_*$ are
the values of $H$ and $\dot\varphi$ at the corresponding time, while if the
curvaton decays before its dominance, it will only contribute part
of the energy density of the universe. Define
$r\equiv\rho_\varphi/\rho_r$, one has: \bea
\zeta^{B}&\simeq&\frac{r}{4}\frac{\delta^{(1)}\rho_\varphi}{\rho_\varphi}
\simeq\frac{r}{4}\frac{\rho_{\varphi,\varphi}}{\rho_\varphi}\delta\varphi
\nonumber\\
\label{zetab}&=&-\frac{\epsilon rH_*}{2\dot\varphi_*}\delta\varphi~,
\eea where the supscripts $A$ and $B$ refer to the two cases
respectively. In both cases, we neglected contribution from
potential term which is subdominant.

The power spectrum of curvature perturbation is defined as $k^3|\zeta|^2/2\pi^2$ in usual convention. Following results from Eqs. (\ref{zetaa}) and (\ref{zetab}), one easily get that for the case $|y|\ll M_p^2$: \bea\label{curvspectrum1}
{\cal P}^{A}_\zeta&\simeq&\frac{27\sqrt{3}H_*^{2}}{4\pi^{2}|y_*|(7-2\epsilon)^{5/2}}\Big(\frac{\epsilon}{3+\epsilon-2\eta}\Big)^{2}~,\\
{\cal
P}^{B}_\zeta&\simeq&\frac{3\sqrt{3}\epsilon^{2}r^{2}H_*^{2}}{16\pi^{2}|y_*|(7-2\epsilon)^{5/2}}~,
\eea where $y_*$ is the value of $y$ at the corresponding time.
The observational data constrains the amplitude of the power
spectrum as ${\cal P}_\zeta=(2.23\pm0.16)\times10^{-9}$ ($68\%$
C.L.), and in usual inflation/curvaton models, to be consistent
with this constraint one needs roughly $H\simeq 10^{-5}M_p$. In
our results we can see, for typical values of parameters
$\epsilon,\eta,r\sim {\cal O}(1)$, we have ${\cal P}_\zeta\sim
H_*^2/|y_*|$, so when $|y|\ll M_p^2$, we may get a lower scale
inflation with $H\ll 10^{-5}M_p$.

Moreover, as one can see from the derivation, the spectral index of the power spectrum of $\zeta$ can be directly inherited from that of $\delta\varphi$, namely Eqs. (\ref{nsphi1}) and (\ref{nsphi2}), without correction during the transfer. Therefore, in order to make our model consistent with observation result $n_s\simeq 0.96$ by PLANCK data \cite{Ade:2013zuv}, one should constrain $\Delta_1$ to be close to about $-0.06$.

In recent years, especially after the release of PLANCK data, the non-Gaussianities of primordial perturbations becomes more and more hot in the studies of the early universe. This is not only due to the great degeneracy in power spectrum of the early universe models, but also because of the more and more accurate measurements of the nonlinear perturbations. In the following of this section, we will focus on the non-Gaussianities generated by our model.

As a curvaton model which the adiabatic perturbations are generated at superhubble scale, the non-Gaussianities are mostly of local type. The local type non-Gaussianities of curvature perturbation are given by:
\be\label{zetang}
\zeta=\zeta_g+\frac{3}{5}f^{local}_{NL}\zeta_g^2~,\ee
where the subscript ``$g$" denotes the Gaussian part of $\zeta$ while $f_{NL}$ is the so-called nonlinear estimator. For local type,
$f^{local}_{NL}$ can be estimated by using the so-called $\delta N$ \cite{Starobinsky:1986fxa}:
\be\label{deltan} \zeta=\delta
N=N_{,\varphi}\delta\varphi+\frac{1}{2}N_{,\varphi\varphi}\delta\varphi^2+...~,
\ee where
$N\equiv\ln a$. Comparing Eqs. (\ref{zetang}) and (\ref{deltan}) one can easily find the relation:
\be
f^{local}_{NL}\Big|_\zeta=\frac{5}{6}\frac{N_{,\varphi\varphi}}{N_{,\varphi}^2}~, \ee and from Eq. (\ref{zeta}), we have: \be N_{,\varphi}=\frac{\rho_{\varphi,\varphi}}{4\rho_r+3(\rho_\varphi+P_\varphi)}~,~N_{,\varphi\varphi}=\frac{\rho_{\varphi,\varphi\varphi}}{4\rho_r+3(\rho_\varphi+P_\varphi)}~,\ee respectively.


Now we consider the two cases separately. For the first case where
curvaton dominates the energy density before decays, one gets:
\bea
f^{local}_{NL}\Big|_\zeta^{A}&\simeq&\frac{5}{2}\frac{(\rho_\varphi+P_\varphi)\rho_{\varphi,\varphi\varphi}}{\rho_{\varphi,\varphi}^2}\nonumber\\
\label{local1}&\simeq&\frac{5}{6}(3+\epsilon-2\eta)~,
\eea
and for the second case where the curvaton decays and never
dominates the energy density, we have
\bea
f^{local}_{NL}\Big|_\zeta^{B}&\simeq&\frac{10}{3r}\frac{\rho_\varphi\rho_{\varphi,\varphi\varphi}}{\rho_{\varphi,\varphi}^2}
\nonumber\\
\label{local2}&\simeq&\frac{5}{r}~,
\eea
respectively. In recent PLANCK paper \cite{Ade:2013tta}, the local-type non-Gaussianities has been constrained as $f_{NL}^{local}=2.7\pm5.8$ ($68\%$ C.L.). With reasonable choices of parameters $\epsilon$, $\eta$ and $r$ to be roughly (or smaller than) ${\cal O}(1)$, one can see that the local-type non-Gaussianities of our model is well within the observational constraints by the PLANCK data.

\section{discussion}
In this paper, we studied a new kind of curvaton model with its
kinetic term nonminimally coupled to the Einstein tensor. This
kind of coupling will contribute a factor of $H^2$ to the kinetic
term of curvaton. Various kinds of the background evolutions of
the curvaton field are reviewed, and a complete analysis of the
perturbation theory of the model, including corrections from
gravitational perturbations, are performed. Thanks to such a
coupling, the perturbations feel like in a nearly de-Sitter
spacetime, which will give rise to scale-invariant power spectrum
favored by the data, independent of the details of the background
evolution of the universe. Although the analysis becomes
complicated when gravitational perturbations are involved in, we
showed that the conclusion still holds qualitatively in
large-speed and small-speed limits. The small tilt of the power
spectrum might be obtained by the corrections from the potential
of the curvaton field. Taking into account the conditions that scalar and tensor perturbations are stable can impose some constraints on the background, but still a quite large range of background EOS could be allowed. Moreover, this simple model can also
generated local-type non-Gaussianities of ${\cal O}(1)$, which is
favored by the recent PLANCK data.

As a natural extention, we note that if $|\epsilon|\gg 1$ is
rapidly changed, the scale factor $a(\eta)$ might evolve as a
constant \cite{slow}. From the last relation of Eq. (\ref{Q}),
$Q\sim 1/t^2$ has to be satisfied. The evolution with
$|\epsilon|\gg 1$ can be parameterized as $H\sim(t_*-t)^{-b}$,
which leads to $Q\sim H^{2/b}$, thus the scale invariance requires
$Q\sim (R/M^2)^{1/b}$. The kinetic term given by such is more
complicated for analytic calculation, and seems hard to be written
in a covariant form as
$G_{\mu\nu}\partial^\mu\varphi\partial^\nu\varphi$. Although
coming from the same logic, this gives us an independent model, so
we leave the discussion on such kind of models for future work.

\section*{Acknowledgments}
T.Q. thanks Keisuke Izumi for useful discussions about the
contents in the appendix. The work of T.Q. are supported by Taiwan
National Science Council under Project No. NSC
101-2923-M-002-006-MY3 and 101-2628-M-002-006- and by Taiwan
National Center for Theoretical Sciences (NCTS). The work of YSP
are supported in part by NSFC under Grant No:11075205, 11222546,
in part by the Scientific Research Fund of GUCAS(NO:055101BM03),
in part by National Basic Research Program of China,
No:2010CB832804.

\appendix
\section{3+1 decomposition}
In deriving perturbed action (\ref{perturb2}) for actions that contains more general gravity terms such as $G_{\mu\nu}\partial^\mu\varphi\partial^\nu\varphi$, it is necessary to know how 3+1 decomposition can be done to such terms. This appendix denotes itself to make clear how the 3+1 form of $G_{\mu\nu}\partial^\mu\varphi\partial^\nu\varphi$ can be obtained. We follow the perturbed metric shown in Eq. (\ref{adm}): \be ds^{2}=-N^{2}dt^{2}+h_{ij}(dx^{i}+N^{i}dt)(dx^{j}+N^{j}dt)~.\ee

First of all, it is useful to define Normal vector of the 3-dimensional hypersurface: $n_{\mu}=n_{0}(dt/dx^{\mu})=(n_{0},0,0,0)$ and $n^{\mu}\equiv g^{\mu\nu}n_{\nu}$. Use the normalization $n_{\mu}n^{\mu}=-1$ one can determine $n_{0}=N$, so \be  n_{\mu}=(N,0,0,0)~,~n^{\mu}=(-\frac{1}{N},\frac{N^{i}}{N})~,\ee and the 3-dimensional induced metric, $H_{\mu\nu}$, which is defined to be orthogonal to the normal vector ($H_{\mu\nu}n^{\nu}=0$), can be chosen as
\be\label{3metric}
H_{\mu\nu}=g_{\mu\nu}+n_{\mu}n_{\nu}~.\ee
Moreover, the corresponding contravariant form can be defined as $H^{\mu\nu}=g^{\mu\nu}+n^{\mu}n^{\nu}$, with $H^{0\mu}=0$.

From now on, one can express $G_{\mu\nu}\partial^\mu\varphi\partial^\nu\varphi$ using 3-metric: \bea &&G_{\mu\nu}\partial^\mu\varphi\partial^\nu\varphi\nonumber\\ &=&G_{\mu\nu}(H^{\mu\alpha}-n^{\mu}n^{\alpha})(H^{\nu\beta}-n^{\nu}n^{\beta})\partial_{\alpha}\varphi\partial_{\beta}\varphi~\nonumber\\ &=&G_{\mu\nu}H^{\mu\alpha}H^{\nu\beta}\partial_{\alpha}\varphi\partial_{\beta}\varphi-2(G_{\mu\nu}n^{\nu}H^{\mu\alpha})\partial_{\alpha}\varphi(n^{\beta}\partial_{\beta}\varphi)\nonumber\\
&&+(G_{\mu\nu}n^{\mu}n^{\nu})(n^{\alpha}\partial_{\alpha}\varphi)(n^{\beta}\partial_{\beta}\varphi)~,\eea however, we still need to express $G_{\mu\nu}H^{\mu\alpha}H^{\nu\beta}$, $G_{\mu\nu}n^{\nu}H^{\mu\alpha}$ and $G_{\mu\nu}n^{\mu}n^{\nu}$ with 1- or 3-dimensional elements in (\ref{adm}). As we will shown below, their expressions are nothing but Gauss, Codazzi and Ricci Equations, which should be familiar to most people who study General Relativity.

Let's first study some properties of the 3-metric, $H_{\mu\nu}$. Firstly, the covariant derivative w.r.t. induced metric $H_{\mu\nu}$ is defined as: \be D_{\mu}\mathcal{T}_{\nu}^{\rho}\equiv H_{\mu}^{\mu^{\prime}}H_{\rho^{\prime}}^{\rho}H_{\nu}^{\nu^{\prime}}\nabla_{\mu}\mathcal{T}_{\nu}^{\rho}~,\ee where $\mathcal{T}_{\nu}^{\rho}$ is an arbitrary tensor, $\nabla$ is the covariant derivative w.r.t. $g_{\mu\nu}$. One can check that $DH_{\mu\nu}=0$. From the property of $H_{\mu\nu}$, one can also have $\overset{(H)}{\Gamma_{jk}^{i}}=\overset{(h)}{\Gamma_{jk}^{i}}$ where they are connections for $H_{\mu\nu}$ and $h_{ij}$ respectively, so one has \be
D_{i}V_{j}=\tilde{\nabla}_{i}V_{j}~,\ee
where $\tilde{\nabla}$ is the covariant derivative w.r.t. $h_{ij}$ ($\tilde{\nabla}h_{ij}=0$.) Moreover, $\overset{(H)}{\Gamma_{\mu\nu}^{0}}=0$ because of the fact that $H^{0\mu}=0$.

Moreover, the curvature of 3-dimensional hypersurface is described by the extrinsic curvature $K_{\mu\nu}$, with the definition:
\bea
K_{\mu\nu}&\equiv&\frac{1}{2}{\cal L}_{n}H_{\mu\nu}\nonumber\\
&=&\frac{1}{2N}(\dot{H}_{\mu\nu}-D_{\mu}N_{\nu}-D_{\nu}N_{\mu})~\eea
where $\mathcal{L}_{n}$ is the Lie derivative w.r.t. $n^{\mu}$. Since $H_{ij}=h_{ij}$ and $D_{i}N_{j}=\tilde{\nabla}_{i}N_{j}$, we have \be K_{ij}=\tilde{K}_{ij}~,\ee the right hand side of which is defined as $K_{ij}=(\dot{h}_{ij}-\tilde{\nabla}_{i}N_{j}-\tilde{\nabla}_{j}N_{i})/2N$. Furthermore, from the relation $K^{\mu\nu}=H^{\mu\mu^{\prime}}H^{\nu\nu^{\prime}}K_{\mu\nu}$ we have $K^{0\mu}=0$ and $K^{ij}=\tilde{K}^{ij}$. Note that $K_{\mu\nu}$ can also be written as
\bea
K_{\mu\nu}&=&H_{\mu}^{\mu^{\prime}}H_{\nu}^{\nu^{\prime}}\nabla_{\mu^{\prime}}n_{\nu^{\prime}}=H_{\mu}^{\mu^{\prime}}\nabla_{\mu^{\prime}}n_{\nu}~,\eea
Therefore it is easy to check that $K_{\mu\nu}n^{\mu}=0$.

The 3-dimensional induced Riemann Tensor (induced means generated by $H_{\mu\nu}$, the same hereafter) is defined by:
\be
^{(3)}R_{\ \mu\nu\rho}^{\sigma}=H_{\sigma^{\prime}}^{\sigma}H_{\mu}^{\mu^{\prime}}H_{\nu}^{\nu^{\prime}}H_{\rho}^{\rho^{\prime}}R_{\ \mu\nu\rho}^{\sigma}-2K_{\nu}^{\ \sigma}K_{\mu\rho}+2K_{\nu\mu}K_{\ \rho}^{\sigma}~,\ee
where $R_{\ \mu\nu\rho}^{\sigma}$ is the 4-dimensional Riemann Tensor (generated by $g_{\mu\nu}$). The indices of $^{(3)}R_{\ \mu\nu\rho}^{\sigma}$ are raised and lowered by $H_{\mu\nu}$. Moreover, it satisfies the relation:
\be
(D_{\mu}D_{\nu}-D_{\nu}D_{\mu})V^{\sigma}=^{(3)}R_{\ \rho\mu\nu}^{\sigma}V^{\rho}\ee
for any spatial vector that satisfies $n_{\sigma}V^{\sigma}=0$.

The contraction of $^{(3)}R_{\ \mu\nu\rho}^{\sigma}$ gives induced Ricci tensor $^{(3)}R_{\mu\nu}$, and from the definition $R_{\mu\nu}\equiv\Gamma_{\mu\nu,\alpha}^{\alpha}-\Gamma_{\mu\alpha,\nu}^{\alpha}+\Gamma_{\mu\nu}^{\alpha}\Gamma_{\alpha\beta}^{\beta}-\Gamma_{\mu\beta}^{\alpha}\Gamma_{\nu\alpha}^{\beta}$  along with the condition $\overset{(H)}{\Gamma_{\mu\nu}^{0}}=0$ and $\overset{(H)}{\Gamma_{jk}^{i}}=\overset{(h)}{\Gamma_{jk}^{i}}$, one can also find that
\be
^{(3)}R_{ij}=\tilde{R}_{ij}~,\ee
where $\tilde{R}_{ij}$ corresponds to $h_{ij}$. Of course by contraction we also have $^{(3)}R=\tilde{R}$.

From this definition of $^{(3)}R_{\ \mu\nu\rho}^{\sigma}$, one can get {\bf the Gauss equation:}
\bea
G_{\mu\nu}n^{\mu}n^{\nu}&=&\frac{1}{2}({}^{(3)}R-K_{\mu\nu}K^{\mu\nu}+K^{2})\nonumber\\
&=&\frac{1}{2}({}^{(3)}\tilde{R}-\tilde{K}{}_{ij}\tilde{K}^{ij}+\tilde{K}^{2})~,\eea
{\bf the Codazzi equation:}
\bea
G_{\mu\nu}n^{\nu}H^{\mu\rho}&=&H^{\rho\sigma}(D_{\alpha}K_{\sigma}^{\alpha}-D_{\sigma}K)\nonumber\\
G_{\mu\nu}n^{\nu}H^{\mu0}&=&0\nonumber\\
G_{\mu\nu}n^{\nu}H^{\mu i}&=&H^{ij}(D_{k}K_{j}^{k}-D_{j}K)\nonumber\\
&=&h^{ij}(\tilde{\nabla}_{k}\tilde{K}_{j}^{k}-\partial_{j}\tilde{K})~,\eea
and {\bf the Ricci equation:}
\begin{widetext}
\bea
G_{\mu\nu}H^{\mu\alpha}H^{\nu\beta}&=&H^{\alpha\gamma}H^{\beta\delta}(\frac{1}{N}{\cal L}_{m}K_{\gamma\delta}-\frac{1}{N}D_{\gamma}D_{\delta}N+^{(3)}R_{\gamma\delta}+KK_{\gamma\delta}-2K_{\gamma\epsilon}K_{\delta}^{\epsilon})\nonumber\\ &&-\frac{1}{2}g_{\mu\nu}H^{\mu\alpha}H^{\nu\beta}[{}^{(3)}R-K^{2}+K_{\rho\sigma}K^{\rho\sigma}+2\nabla_{\rho}(Kn^{\rho}-n^{\sigma}\nabla_{\sigma}n^{\rho})]\ (m^{\mu}=Nn^{\mu})\\
G_{\mu\nu}H^{\mu0}H^{\nu\beta}&=&G_{\mu\nu}H^{\mu\alpha}H^{\nu0}=0~,\\
G_{\mu\nu}H^{\mu i}H^{\nu j}&=&\frac{1}{N}\dot{\tilde{K}}_{ij}-\frac{1}{N}(N^{k}\tilde{\nabla}_{k}\tilde{K}_{ij}+\tilde{K}^{jk}\tilde{\nabla}_{i}N_{k}+\tilde{K}^{ik}\tilde{\nabla}_{j}N_{k})-\frac{1}{N}\tilde{\nabla}_{i}(\partial_{j}N)+^{(3)}\tilde{R}_{ij}+\tilde{K}\tilde{K}_{ij}-2\tilde{K}_{ik}\tilde{K}_{j}^{k}\nonumber\\ &&-\frac{1}{2}h^{ij}[{}^{(3)}\tilde{R}-\tilde{K}^{2}+\tilde{K}_{kl}\tilde{K}^{kl}+\frac{2}{N\sqrt{h}}\partial_{0}(\sqrt{h}\tilde{K})-\frac{2}{N\sqrt{h}}\partial_{k}(\sqrt{h}N^{k}\tilde{K}+\sqrt{h}\partial^{k}N)]~.\eea
\end{widetext}
These three equations shows the (3+1)-form of the Einstein tensor (or equivalently, Ricci tensor). Moreover, by contraction we have:
\bea\label{ricci}
R&=&^{(3)}R-K^{2}+K_{\mu\nu}K^{\mu\nu}+2\nabla_{\mu}(Kn^{\mu}-n^{\nu}\nabla_{\nu}n^{\mu})~\nonumber\\
&=&^{(3)}\tilde{R}-\tilde{K}^{2}+\tilde{K}_{ij}\tilde{K}^{ij}+\frac{2}{N\sqrt{h}}\partial_{0}(\sqrt{h}\tilde{K})~\nonumber\\ &&-\frac{2}{N\sqrt{h}}\partial_{i}(\sqrt{h}N^{i}\partial_{i}\tilde{K}+\sqrt{h}\partial^{i}N)~\eea
for Ricci scalar. Till now, all the needed variables of Gravity part have been decomposed and presented in terms of $N$, $N_{i}$ and $h_{ij}$-related variables, which become computable. We refer the readers to \cite{Wald:1984rg} for more complete arguments.



\end{document}